\documentstyle[epsfig,rotating,epsf,12pt,here]{article} 
\def\Title#1{\begin{center} {\Large {\bf #1} } \end{center}}

\begin{document}

\Title{Accelerator and Reactor Neutrino Experiments}

\bigskip\bigskip
\begin{raggedright}  

{\it L. DiLella\index{DiLella, L.}\\
CERN, Geneva, Switzerland}
\bigskip\bigskip
\end{raggedright}

\section{Introduction}

Since the first detection of the neutrino at reactors \cite{ref1} and the discovery of the $\nu_\mu$ at
the Brookhaven AGS \cite{ref2}, accelerators and reactors have been used to provide intense beams of
neutrinos. Over the last 30 years, experiments with neutrino beams have played a major role in
shaping up the Standard Model of particle physics as it is known today. 

At present, electroweak theory is more conveniently studied at high energy hadron and electron colliders
where the $W$ and $Z$ bosons are directly produced and their properties can be studied in great detail.
 Nevertheless, experiments with neutrino beams
continue to play a major role in particle physics by addressing other crucial questions: are neutrino
massive? do they mix?

The search for neutrino oscillations is presently the most promising method to search for very small
neutrino masses. The hypothesis of neutrino mixing
 postulates that the three known neutrino flavors, $\nu_e,~\nu_\mu$ and
$\nu_\tau$, are not mass eigenstates but quantum-mechanical superpositions of three mass
eigenstates, $\nu_1,~\nu_2$ and
$\nu_3$, with mass eigenvalues $m_1,~m_2$ and $m_3$, respectively:
\begin{equation}
\nu_\alpha = \sum_k~U_{\alpha k}~\nu_{k}
\end{equation}

In Eq. (1) $\alpha = e, \mu,\tau$ is the flavor index, $k = 1,2,3$ is the index of the mass
eigenstates and $U$ is a unitary 3 $\times$ 3 matrix.  
If mixing of two neutrino is dominant, the
probability to detect $\nu_\beta$ if the neutrino state at production is pure $\nu_\alpha$
can be written as
\begin{equation}
P_{\alpha\beta}(L) = \sin^2(2\theta)~\sin^2\Biggl(1.27~\Delta m^2~\frac{L}{E}\Biggl)
\end{equation}
where $\theta$ is the mixing angle, $L$ is the distance between the neutrino source and the detector in
Km, $\Delta m^2 = |m^2_1 - m^2_2|$ is in eV$^2$ and $E$ is in GeV. Depending on $L$ and $E$,
experiments are sensitive to different $\Delta m^2$ regions. 

\section{Neutrino oscillation searches at nuclear reactors}

Nuclear reactors are intense, isotropic sources of $\bar{\nu}_e$ produced by $\beta$-decay of fission
fragments. The $\bar{\nu}_e$ energy is below 10 MeV, with an average value of $\sim$ 3 MeV. Their energy
spectrum and flux above 2 MeV have been measured \cite{ref3}. From the knowledge of the reactor
parameters the $\bar{\nu}_e$ flux is predicted with an uncertainty of 2.7\%.

The inverse $\beta$-decay reaction
\begin{equation}
\bar{\nu}_e + p \to e^+ + n
\end{equation}
is used to detect reactor neutrinos. This reaction has a threshold of 1.8 MeV and gives rise to two
detectable signals: a prompt one from $e^+$ production and annihilation into two photons; and a late
photon signal from neutron capture (2.2 MeV from the reaction $np \to d\gamma,~\sim$ 8 MeV from neutron
capture by Gadolinium in Gd-doped scintillator).

Table 1 lists the main parameters of the two long baseline experiments which have recently reported
results.

The Chooz experiment is named after the site of a nuclear power plant (the village of Chooz in the
Ardennes region of France). The measured positron spectrum agrees with the spectrum expected in the
absence of neutrino oscillations, as shown in Fig. 1. The energy-integrated ratio between the
measured and expected event rate is found to be \cite{ref4}
\begin{equation}
R = 1.010 \pm 0.028~({\rm stat.}) \pm 0.027~({\rm syst.})
\end{equation}

\begin{table}[H]
\begin{center}
\caption{Long baseline reactor experiments.}
\begin{tabular}{|l|c|c|}  
\hline
&Chooz \cite{ref4}&Palo Verde \cite{ref5}\\
\hline
\multicolumn{1}{|c|}{Number of reactors}&2&3\\
\hline
Distance (Km)&1.114, 0.998&0.89, 0.89, 0.75\\
\hline
Thermal power (GW)&8.5&11.0\\
\hline
Detector type&Gd-doped&Gd-doped\\
&homogeneous&segmented\\
&scintillator&scintillator\\
\hline
Detector mass&5 Tons& 12 Tons\\
\hline
Rock overburden&300 m w.e.&32 m w.e.\\
\hline
Event rate (at full power)&25.5 $\pm$ 1.0~d$^{-1}$&39.1 $\pm$ 1.0~d$^{-1}$\\
\hline
Event rate (reactors off)&1.1 $\pm$ 0.3~d$^{-1}$&32.6 $\pm$ 1.0~d$^{-1}$\\
\hline
Status&Completed&In progress\\
\hline
\end{tabular}
\end{center}
\end{table}

Figure 2 shows the region of $\bar{\nu}_e - \bar{\nu}_x$ oscillation parameters excluded at the 90\%
confidence level. This result strongly constrains the contribution from a possible $\nu_\mu - \nu_e$
oscillation to the deficit of atmospheric $\nu_\mu$ observed by many experiments \cite{ref6}.

\begin{figure}[H]
\centering\epsfig{figure=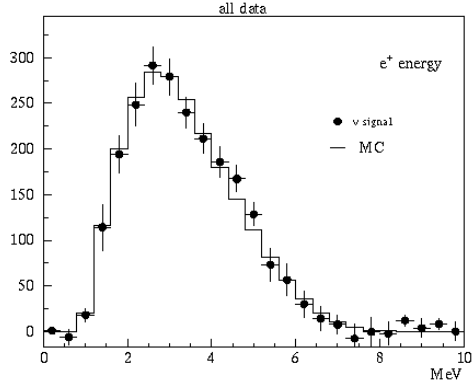,width=6.6cm}
\caption[]{Expected positron spectrum for the case of no oscillation (histogram), superimposed on the
measured spectrum in the
Chooz experiment \cite{ref4}.}
\end{figure}

\begin{figure}[H]
\centering\epsfig{figure=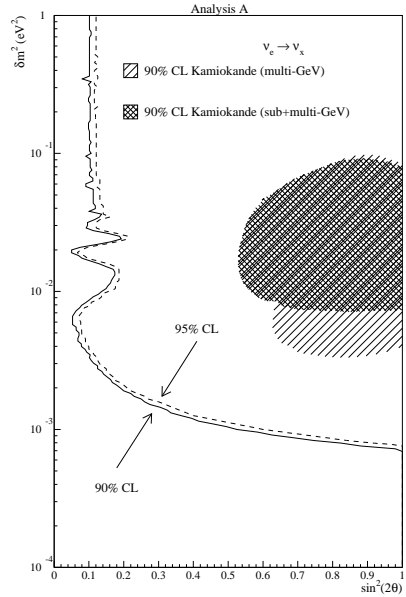,width=6.0cm}
\caption[]{Boundary of the $\bar{\nu}_e-\bar{\nu}_x$ oscillation parameter region excluded by the
comparison of the measured positron spectrum with the spectrum expected in the absence of $\bar{\nu}_e$
oscillations \cite{ref4}. Also shown is the region of $\nu_\mu - \nu_e$ oscillation parameters
allowed by the results of Ref. \cite{ref7}.}
\end{figure}

An analysis of the Chooz data that does not rely on the absolute knowledge of the $\bar{\nu}_e$ flux has
also been performed by taking advantage of the fact that a substantial amount of data was taken with only
one reactor in operation. The comparison of the measured positron spectra from each reactor
provides a direct determination of the oscillation probability because, as shown in Table 1, the two
reactors are at different distances. The region of  oscillation parameters excluded by this analysis
\cite{ref4} is shown in Fig. 3.

\begin{figure}[H]
\centering\epsfig{figure=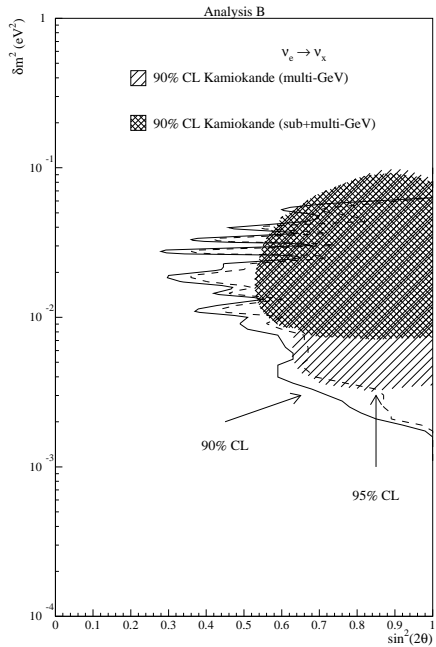,width=6.5cm}
\caption[]{Exclusion contours at 90\% and 95\% confidence level obtained from the
ratio of the positron spectra from the two reactors, as measured by the Chooz
experiment. Also shown is the region of
$\nu_\mu-\nu_e$ oscillation parameters allowed by the results of Ref. \cite{ref7}.}
\end{figure}

In the Palo Verde experiment \cite{ref5} because of the
shallow site the background is much higher than in the Chooz experiment (see Table 1). 
 From the data collected in 1998 during the first 70 day run this experiment has reported
the value
$$R = 1.3 \pm 0.3~({\rm stat.)} \pm 0.2~({\rm syst.})$$
for the ratio between the measured event rate and the rate expected in the absence of $\bar{\nu}_e$
oscillations. This result is much less sensitive to oscillations than the final result from the Chooz
experiment (see Eq. 4).

\section{Future oscillation searches at reactors}

KAMLAND \cite{ref8} is a long baseline reactor experiment which will begin data taking in 2001. The
detector consists  of $\sim$ 1000 tons of  scintillating isoparaffin oil,  installed in the Kamioka
mine at a depth of 2700 m of water equivalent.

KAMLAND aims at detecting the $\bar{\nu}_e$ produced by five nuclear reactors
located at distances between 150 and 210 km from the detector and producing a
total thermal power of 127 GW. When all reactors run at full power, the event rate is expected to be
$\sim$ 2 d$^{-1}$ with a signal-to-noise ratio of 10. The background is measured by observing the
variation of the event rate with the reactor power.

Because of its large distance from the reactors, KAMLAND is sensitive to $\Delta m^2 > 7 \times
10^{-6}~{\rm eV}^2$ and sin$^2 2\theta > 0.1$, a region which includes the large mixing angle, large
$\Delta m^2$ MSW solution of the solar neutrino problem \cite{ref9}. The anticipated exclusion region
after three years of data taking in the absence of neutrino oscillations is shown in Fig. 4.

\begin{figure}[H]
\centering\epsfig{figure=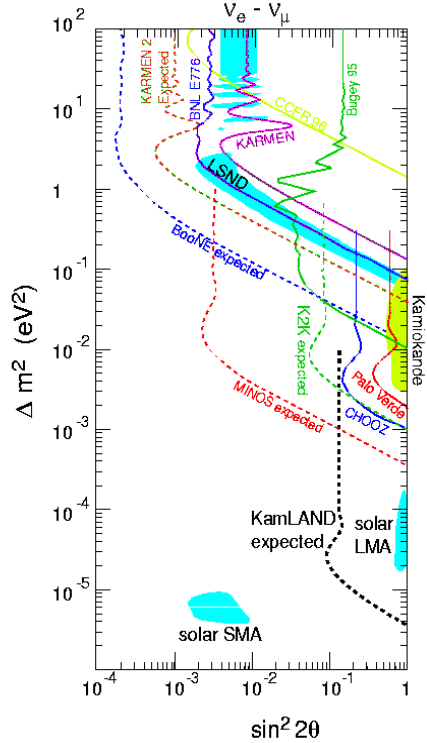,width=5.5cm}
\caption[]{Region of $\bar{\nu}_e - \bar{\nu}_x$ oscillation parameters excluded at the 90\%
confidence level if no oscillation signal is detected by KAMLAND after three years of data
taking. Also shown are regions of oscillation parameters allowed or excluded by other
experiments. The solar large mixing (LMA) and small mixing angle (SMA) MSW solutions are also
visible.}
\end{figure}

\section{Searches for $\nu_\mu - \nu_e$ oscillations at accelerators}

The Liquid Scintillator Neutrino Detector (LSND) \cite{ref10} and the
KArlsruhe-\break\hfill Rutherford Medium Energy Neutrino (KARMEN) experiment \cite{ref11} use
neutrinos produced in the beam stop of a proton accelerator. LSND has finished data taking
at the Los Alamos Neutron Science Center (LANSCE) at the end of 1998, while KARMEN is still
running at the ISIS neutron spallation facility of the Rutherford-Appleton Laboratory.

In these experiments neutrinos are produced by the following decay processes :
\begin{itemize}
\item[(i)] $\pi^+\to\mu^+\nu_\mu$ (in flight or at rest);
\item[(ii)] $\mu^+ \to \bar{\nu}_\mu e^+\nu_e$ (at rest);
\item[(iii)]$\pi^-\to\mu^-\bar{\nu}_\mu$ (in flight);
\item[(iv)] $\mu^-\to \nu_\mu e^-\bar{\nu}_e$ (at rest).
\end{itemize}

The  $\bar{\nu}_e$ yield is very small (of the order of $4 \times 10^{-4}$ with respect to
$\bar{\nu}_\mu$) because
$\pi^-$ decaying in flight are a few \% of all produced $\pi^-$ and only a small fraction of $\mu^-$ stopping in heavy
materials decays to $\nu_\mu e^-\bar{\nu}_e~(\pi^-$ at rest are immediately captures by nuclei; most $\mu^-$ stopping
in high-$Z$ materials undergo the capture process $\mu^- p \to \nu_\mu n$). 

Table 2 lists the main parameters of the two experiments. For the $\bar{\nu}_\mu -
\bar{\nu}_e$ oscillation search, the
$\bar{\nu}_e$ is detected by   reaction (3). The
delayed neutron signal  results from the 2.2 MeV $\gamma$-ray from the reaction $np \to d\gamma$ and
also, for KARMEN, the 8 MeV line from $\gamma$-rays
emitted by neutron capture in Gadolinium, which is contained in thin layers of $Gd_2O_3$
placed between adjacent cells.

While the LANSCE beam is ejected in $\sim 500 \mu$s long spills 8.3 ms apart, the ISIS beam is pulsed with a
time structure consisting of two 100 ns long pulses separated by 320 ns (this sequence has a repetition rate of 50 Hz).
Thus it is possible to separate neutrinos from muon and pion decay from their different time distributions with respect
to the beam pulse.

Table 3 lists preliminary results from LSND \cite{ref12} and KARMEN \cite{ref13}, obtained
after requiring space and time correlation between the prompt and delayed signal, as
expected from $\bar{\nu}_e p \to e^+n$ (and for KARMEN requiring also the time correlation
between the $e^+$ signal and the beam pulse). The LSND result gives evidence for an excess of
$\bar{\nu}_e$ events with a statistical significance of $\sim$ 4.5 standard deviations.

\begin{center}
\begin{table}[H]
\caption[]{Parameters of the LSND and KARMEN experiments.}
\vskip0.2cm
\begin{tabular}{|l|c|c|}
\hline
&LSND&KARMEN\\
\hline
Proton   Kin. Energy&800 MeV&800 MeV\\
\hline
Proton beam current&1000 $\mu$A&200~$\mu$A\\
\hline
&Single cylindrical&512 cells filled\\
&tank; 1220 PMT's;&with liquid scintillator;\\
Detector&collection of both&cell dim. 18 $\times~18\times~350$ cm\\
&scintillation and&\\
&$\breve{\rm C}$erenkov light.&\\
\hline
Detector mass&167 tons&56 tons\\
\hline
Event localization&timing&cell size\\
\hline
Distance from&29 m&17 m\\
$\nu$ source&&\\
\hline
Angle between&&\\
$\nu$ direction and&17$^\circ$&90$^\circ$\\
proton beam&&\\
\hline
Data taking period&1993--98&Feb. 97--Feb. 99\\
\hline
Protons on target&1.8 $\times~10^{23}$&2.9 $\times~10^{22}$\\
\hline
\end{tabular}
\end{table}
\end{center}

\begin{center}
\begin{table}[H]
\caption[]{Preliminary results from LSND and KARMEN.}
\vskip0.2cm
\footnotesize{\begin{tabular}{|l|c|c|}
\hline
&LSND [12]&KARMEN [13]\\
\hline
$e^+$ energy interval&20--60 MeV& 16--50 MeV\\
\hline
Observed events&70&8\\
\hline
Cosmic ray background&17.7 $\pm$ 1.0&1.9 $\pm$ 0.1\\
\hline
Total background&30.5 $\pm$ 2.7& 7.82 $\pm$ 0.46\\
\hline
$\bar{\nu}_e$ signal  events& 39.5 $\pm$ 8.8&$<$ 3.4 (90\% C.L)\\
\hline
$\bar{\nu}_\mu-\bar{\nu}_e$ oscillation probability&(3.3 $\pm~0.9\pm0.5)
 \times 10^{-3}$&$<1.03 \times 10^{-3}$~(90\% C.L.) \\
\hline
\end{tabular}}
\end{table}
\end{center}

Figure 5 shows the preliminary $e^+$ energy distribution of the 70 events
observed by LSND, together with the distributions expected from
backgrounds and from $\bar{\nu}_\mu -\bar{\nu}_e$ oscillations for
two different $\Delta m^2$ values. The region of oscillation
parameters describing the LSND result is shown in Fig. 6, together
with the region excluded by the Bugey reactor experiment \cite{ref14}
and by the present 
KARMEN result. This figure shows that, if the LSND result is
correct, the only allowed region of $\bar{\nu_\mu}-\bar{\nu}_e$
oscillation parameters is a narrow strip with $\Delta m^2$ between
0.2 and 2 eV$^2$ and sin$^2 2\theta$ between 0.002 and 0.04.

\begin{figure}[H]
\centering\epsfig{figure=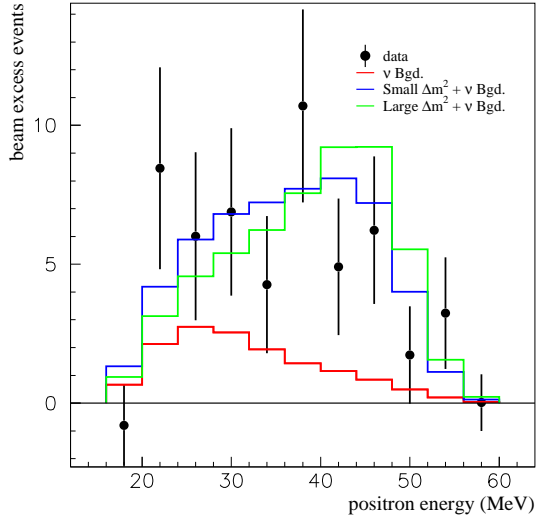,width=7cm}
\caption[]{Preliminary $e^+$ energy distribution of the 70 events observed by LSND. Also shown are
the distributions expected from backgrounds (histogram with error bars) and the expectations from
$\bar{\nu}_\mu - \bar{\nu}_e$ oscillations for two different $\Delta m^2$ values.}
\end{figure}

\begin{figure}[H]
\centering\epsfig{figure=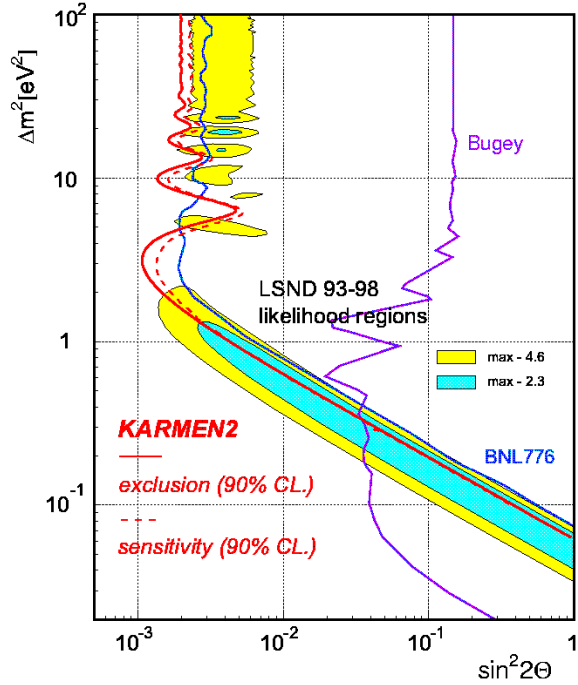,width=7.5cm}
\caption[]{Region of $\bar{\nu}_\mu - \bar{\nu}_e$ oscillation parameters allowed by the preliminary
LSND results \cite{ref12}. Also shown are the boundaries of the regions excluded by the Bugey reactor experiment \cite{ref14}
and by the recent KARMEN result \cite{ref13}.}
\end{figure}

During the first three years of LSND data taking the target area of
the LANSCE accelerator consisted of a 30 cm long water target
located $\sim$ 1 m upstream of the beam stop. This configuration
enhanced the probability of pion decay in flight, allowing LSND to
search for $\nu_\mu-\nu_e$ oscillations using $\nu_\mu$ with energy
above 60 MeV. In this case one expects to observe an
excess of events from the reaction
$$\nu_e + C^{12} \to e^- + X$$
above the expected backgrounds. This reaction has only one
signature (a prompt signal) but the higher energy, the longer track
and the directionality of $\breve{\rm C}$erenkov light help improving
electron identification and measuring its direction.

In this search \cite{ref15} LSND has observed 40 events to be compared
with\break\hfill 12.3 $\pm$ 0.9 events from cosmic ray background and 9.6 $\pm$
1.9 events from machine-related (neutrino-induced) processes. The
excess of events  (18.1 $\pm$ 6.6 events) corresponds to a $\nu_\mu
-\nu_e$ oscillation probability of $(2.6 \pm 1.0) \times 10^{-3}$,
consistent with the value found from the study of the $\bar{\nu}_e
p \to e^+ n$ reaction below 60 MeV.

In view of the importance of the LSND result, it is useful to review other measurements of
conventional neutrino processes in the LSND experiment and to compare them with theoretical
predictions
\cite{ref12}. Such a comparison is presented in Table 6. There is general agreement between the
measured and predicted values, except for the last reaction in Table 4. However, in this case the
experimental value for the cross-section is lower than the predicted one and the theoretical
calculation is difficult because of nuclear effects.

\begin{table}[H]
\begin{center}
\caption[]{Cross-sections for conventional neutrino processes: theoretical predictions and LSND
measurements \cite{ref12}.}
\begin{tabular}{|l|c|c|c|}  
\hline
\multicolumn{1}{|c|}{Reaction}&$\sigma_{theory}$~(cm$^2$)&\multicolumn{2}{|c|}{LSND
measurement}\\\cline{2-4}
&&Events above&$\sigma_{meas}$~(cm$^2$)\\
&&background&\\
\hline
$\nu_e + C^{12}\to e^-+N^{12}$&9.3 $\times~10^{-42}$&515&$(9.1 \pm 0.4 \pm 0.9) \times 10^{-42}$\\
$\nu_e + C^{12}\to e^-+(N^{12})^*$&6.3 $\times~10^{-42}$&660&$(5.7 \pm 0.6 \pm 0.6) \times 10^{-42}$\\
$\nu_\mu + C^{12}\to \mu^-+N^{12}$&6.4 $\times~10^{-41}$&57&$(6.6 \pm 1.0 \pm 1.0) \times 10^{-41}$\\
$\nu_\mu + C^{12}\to \mu^-+(N^{12})^*$&20.5 $\times~10^{-40}$&1738&$(11.2 \pm 0.3 \pm 1.8) \times
10^{-40}$\\
\hline
\end{tabular}
\end{center}
\end{table}

\subsection{Future searches for $\nu_\mu-\nu_e$ oscillations at accelerators}

The KARMEN experiment will finish data taking in the year 2001. By then, 
its sensitivity to
$\bar{\nu}_\mu-\bar{\nu}_e$ oscillations will
have improved by a factor of 1.7 with respect to the present value.
However, in case of a negative result the exclusion region will not
fully contain the region allowed by LSND. A new search for
$\nu_\mu-\nu_e$ (or $\bar{\nu}_\mu - \bar{\nu}_e)$ oscillations is
needed, therefore, to unambiguously confirm or refute the LSND signal.

The Mini-BOONE experiment \cite{ref16} is a first phase of a new,
high sensitivity search for $\nu_\mu-\nu_e$ oscillations (BOONE is
the acronym for Booster Neutrino Experiment). Neutrinos are
produced using an 8 GeV, high intensity proton beam from the
Fermilab Booster Synchrotron. The beam consists mainly of $\nu_\mu$
from $\pi^+$ decay with a small contamination $(\sim$ 0.3\%) of
$\nu_e$ and a broad energy distribution from 0.3 to 2 GeV.

The Mini-BOONE detector will be installed at a distance of 500 m from
the neutrino source. It consists of a 6 m radius spherical tank
filled with mineral oil. $\breve{\rm C}$erenkov light produced in the oil
is collected by $\sim$ 1500 photomultiplier tubes located on the
surface of the sphere. The detector is surrounded by
anticoincidence counters and will use the different pattern of
$\breve{\rm C}$erenkov light expected for muons, electrons and $\pi^0$ to
identify these particles. For a fiducial mass of 445 tons Mini-BOONE expects to detect $\sim
5 \times 10^5~\nu_\mu~C^{12}\to \mu^- X$ events and $\sim$ 1700
$\nu_eC^{12} \to e^- X$ events in one year. A $\nu_\mu-\nu_e$ oscillation with the parameters
required to describe the LSND signal would result in an excess of $\sim$ 1500 $\nu_e~C^{12} \to e^-
X$ events. 

If no oscillation signal is observed, Mini-BOONE will exclude the region of oscillation
parameters shown in Fig. 7. However, if a signal is
observed, it should be possible to measure precisely the
oscillation parameters  by using a second detector at a
different distance.

Mini-BOONE will begin taking data in the year 2002.

\begin{figure}[H]
\centering\epsfig{figure=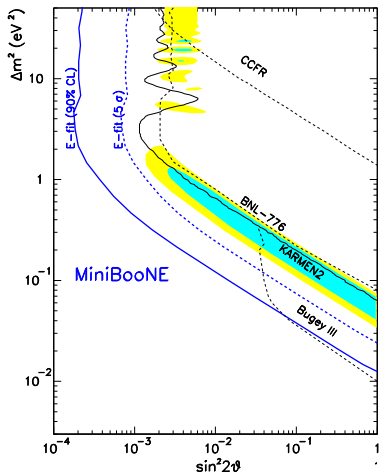,width=5cm}
\caption[]{Region of $\nu_\mu-\nu_e$ oscillation parameters excluded by Mini-BOONE if no signal is
observed after one year of data taking.}
\end{figure}

\subsection{Searches for $\nu_\mu-\nu_\tau$  oscillations}

Two experiments searching for $\nu_\mu-\nu_\tau$ oscillations have
recently completed data taking at CERN. They both used the
wide-band neutrino beam from the CERN 450~GeV proton synchrotron
(SPS). The method adopted by both experiments consists in detecting
$\tau^-$ production with a sensitivity corresponding to a
$\nu_\tau/\nu_\mu$ ratio much larger than the value expected from
conventional $\nu_\tau$ sources in the beam (the main $\nu_\tau$
production process is $D_s$ production by the primary protons,
followed by the decay $D_s \to \tau\nu_\tau$). The observation of
$\tau^-$ could only result, therefore, from $\nu_\mu-\nu_\tau$
oscillations.

The two experiments, CHORUS and NOMAD, are installed one behind the
other at a distance of $\sim$ 820 m from the proton target.  The distance
between the proton target and the end of the decay tunnel is 414 m.
Table 5
lists the mean neutrino energies and relative abundances. The
$\nu_\tau$ `natural' abundance is estimated to be $\sim 5 \times
10^{-6}$~\cite{ref17}.

\begin{center}
\begin{table}[H]
\caption[]{Mean energies and relative abundances for $\nu$ fluxes and
CC interactions at the NOMAD detector. The integrated
$\nu_\mu$ flux is $1.11 \times 10^{-2}~\nu_\mu$ per proton on
target}
\vskip0.2cm
\begin{tabular}{|c|c|c|c|c|}
\hline
&\multicolumn{2}{|c|}{Flux}&\multicolumn{2}{|c|}{Charged-current
interactions}\\
\hline
$\nu$ type&$<E_\nu>$&Relative&$<E_\nu>$&Relative\\
&GeV&abundance&GeV&abundance\\
\hline
$\nu_\mu$&23.5&1.00&42.6&1.00\\
$\bar{\nu}_\mu$&19.2&0.061&41.0&0.0249\\
$\nu_e$&37.1&0.0094&56.7&0.0148\\
$\bar{\nu}_e$&31.3&0.0024&53.6&0.0016\\
$\nu_\tau$&$\sim$ 35&$\sim 5 \times 10^{-6}$&&\\
\hline
\end{tabular}
\end{table}
\end{center}

CHORUS  (Cern Hybrid Oscillation Research apparatUS) aims at detecting the decay of
the short-lived $\tau$ lepton in nuclear emulsion. This technique provides a space
resolution of $\sim~1~\mu$m, well matched to the average $\tau^-$ decay length of 1
mm.

The apparatus \cite{ref18} consists of an emulsion target
with a total mass of\break\hfill $\sim$ 770 kg, followed by an electronic tracking detector
made of scintillating fibres, an aircore hexagonal magnet, electromagnetic and
hadronic calorimeters and a muon spectrometer consisting of magnetized iron toroids
interleaved with drift chambers. Neutrino events with a $\mu^-$ or a negatively charged hadron are
selected and the tracks are followed back to the exit point from the emulsion target. The method
is illustrated in Fig. 8. It relies on special emulsion sheets mounted between the target
and the fibre tracker (these sheets are replaced every few weeks during the run). With the
reconstruction accuracy of the fibre tracker the track position  on the special sheet is
predicted  within an area of 360 $\mu$m $\times~360~\mu$m. In this area, because of
the short exposure time, one finds, an average, 5 muon tracks which are rejected by
angular measurement. The search is then continued in an area of 20~$\mu$m
$\times~20~\mu$m into the emulsion target, with negligible background despite the
long exposure time of the target (2 years).

\begin{figure}[H]
\centering\epsfig{figure=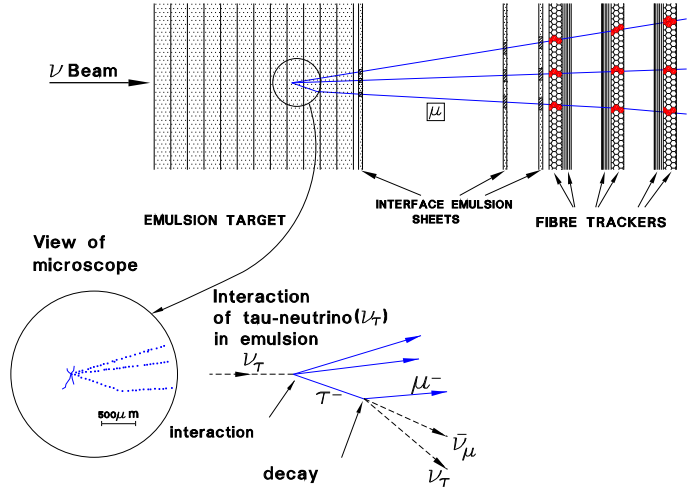,width=9cm}
\caption[]{Expected configuration of a typical $\nu_\tau$ charged-current interaction in the CHORUS
emulsion. In this example the $\tau^-$ decays to $\mu^- \nu_\tau
\bar{\nu}_\mu$.}
\end{figure}

Events in which the neutrino interaction point is found in the emulsion target are
analysed to search for $\tau^-$ decay, which is identified by the presence of a
change in direction (kink) of a negatively charged track, as expected from
one-prong decays. No other charged leptons must be observed at the primary vertex
and the transverse momentum $(p_T$) of the selected particle with respect to the
$\tau$ candidate direction is required to be larger than 0.25 GeV (to eliminate
strange particle decays). Negative tracks with impact parameters larger than
$\sim~8~\mu$m are also considered as possible secondary particles from $\tau^-$
decay.

Negative muons with momentum $p <$ 30 GeV or negative hadrons with $1 < p$\break\hfill $<$ 20 GeV
are considered as possible $\tau^-$ decay products. If observed, the kink is required to be
within 3.95 mm from the interaction point for muon tracks, and 2.37 mm for hadron
tracks. The emulsion scanning procedure to locate the interaction point and to
reject events incompatible with $\tau^-$ decays is fully automatic. The remaining
events undergo a computer assisted eye scan to confirm the presence of a $\tau$
decay. 

CHORUS took data between May 1994 and  the end of 1997. Table
6 summarizes the present status of the analysis \cite{ref19}. No $\tau^-$
candidate has been observed yet.

The dominant background in the one-muon channel is the production of charmed
particles from $\bar{\nu}_\mu$ interactions
$$\bar{\nu}_\mu~ N \to \mu^+D^-X$$
followed by the decay $D^- \to \mu^-$ + neutral particles. These events are
dangerous only if the $\mu^+$ is not identified. Their contribution to the data
sample analysed so far is 0.24 $\pm$ 0.05 events. In the muonless channel the background from charm
production amounts to 0.075 $\pm$ 0.015 events. A more serious background is the
interaction of negative hadrons with nuclei producing only one outgoing negatively charged particle 
with no evidence for nuclear break-up (these interactions are called `white kinks'). The rate of
such events, of the order of one event, is affected by a large uncertainty.

\begin{center}
\begin{table}[H]
\caption[]{Status of the CHORUS analysis \cite{ref19}}
\vskip0.2cm
\begin{tabular}{|l|c|c|}
\hline
&One-muon events&Muonless events\\
Expected number of events&458.6 $\times~10^3$&116 $\times~10^3$\\
\hline
Fraction scanned so far&54\%&47\%\\
\hline
Events with identified $\nu$ interaction&126.2 $\times~10^3$&19.4 $\times~10^3$\\
\hline
$N_\tau$ for $<P_{\mu\tau}>$ = 1&4876&1137\\
\hline
\end{tabular}
\end{table}
\end{center}

The expected numbers of events, $N_\tau$, for an oscillation probability $P_{\mu\tau}$
= 1, are given in the last row of Table 8. The 90\% confidence upper limit on
$P_{\mu\tau}$ is
$$P_{\mu\tau} <  4.0 \times 10^{-4}~.$$
 The region of oscillation parameters excluded by this
result under the assumption of two-neutrino mixing is shown in Fig. 9 together
with the results from previous experiments \cite{ref20}. A more stringent limit, $P_{\mu\tau} <2.6 \times 10^{-4}$, would be obtained \cite{ref21}
using the method recently proposed by Feldman and Cousins \cite{ref22}.

\begin{figure}[H]
\centering\epsfig{figure=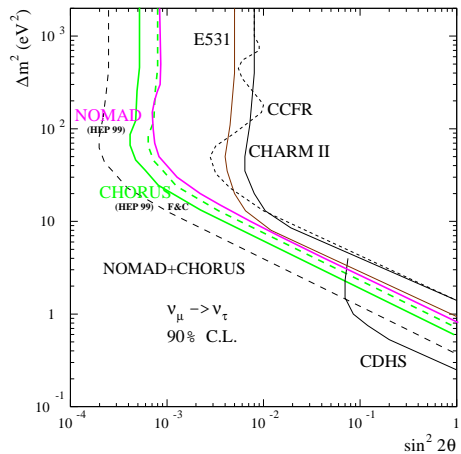,width=6cm}
\caption[]{The $\Delta m^2-\sin^2(2\theta)$ plane for $\nu_\mu - \nu_\tau$ oscillation.The regions
excluded by CHORUS and NOMAD, and by their combined results \cite{ref21}, are shown
together with the results of previous experiments \cite{ref20}. Full lines: CHORUS
(for two different statistical methods); dashed lines: NOMAD and combined result.}
\end{figure}

CHORUS is expected to reach the upper limit $P_{\mu\tau} < 10^{-4}$ is no event
is found after the completion of the analysis.

NOMAD (Neutrino Oscillation MAgnetic Detector) is designed to search for $\nu_\mu
- \nu_\tau$ oscillations by observing $\tau^-$ production using kinematical
criteria \cite{ref23}, which require a precise measurement of secondary particle
momenta. The main detector components are \cite{ref24}  (see Fig. 10):
\begin{itemize}
\item[--] drift chambers (DC) used to reconstruct charged particle tracks and
also acting as the neutrino target (fiducial mass $\sim$ 2.7 tons, average
density 0.1 g/cm$^3$, radiation length $\sim$ 5 m);
\item[--] nine independent transition radiation detectors (TRD) for electron
identification, interleaved with additional drift chambers;
\item[--]  an electromagnetic calorimeter (ECAL) located behind a `preshower'
detector (PRS);
\item[--] a hadronic calorimeter (HCAL);
\item[--] large-area muon chambers.
\end{itemize}

\begin{figure}[H]
\centering\epsfig{figure=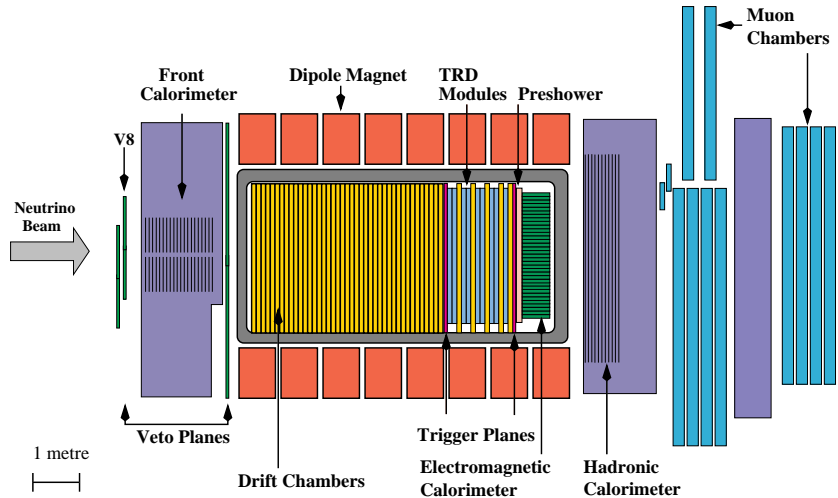,width=11cm}
\caption[]{Side view of the NOMAD detector.}
\end{figure}

DC, TRD and ECAL are located inside a uniform magnetic field of 0.4~T
perpendicular to the beam direction.

The NOMAD experiment aims at detecting $\tau^-$ production by observing both
leptonic and hadronic decay modes of the $\tau^-$. The decay $\tau^-\to\nu_\tau
e^-\bar{\nu}_e$ is particularly attractive because the main background results
from $\nu_e$ charged-current interactions which are only $\sim$ 1.5\% of the
total number of neutrino interactions in the target fiducial volume and have an
energy spectrum quite different from that expected from $\nu_\mu - \nu_\tau$
oscillations (see   Table 5). The selection of this decay relies on the
presence of an isolated electron in the final state and on the correlation among
the lepton transverse momentum $(\vec{p}^e_T)$, the total transverse momentum of
the hadronic system $(\vec{p}^H_T)$ and the missing transverse momentum
$(\vec{p}^m_T)$ (only the momentum components perpendicular to the beam direction
can be used because the incident neutrino energy is unknown). In the case of
$\nu_e$ charged-current events $\vec{p}^e_T$ is generally opposite to
$\vec{p}^H_T$ and $|\vec{p}^m_T|$ is small (it should be exactly zero if the
momenta of all secondary particles were measured precisely and the target
nucleon were at rest). On the contrary, in $\tau^- \to \nu_\tau e^-\bar{\nu}_e$
decays there is a sizeable $|\vec{p}^m_T |$ associated with the two outgoing
neutrinos. Furthermore, in a large fraction of events $\vec{p}^m_T$ is at
opposite azimuthal angles to $\vec{p}^H_T$, in contrast with $\nu_e$
charged-current interactions for which large values of $|\vec{p}^m_T|$ result
mostly from hadrons escaping detection (in these cases the azimuthal separation
between $\vec{p}^H_T$ and $\vec{p}^m_T$ is small).

In practice, for most $\tau^-$ decay channels the method to separate the signal
from backgrounds uses ratios of likelihood functions. These functions are
approximated by products of probability density functions of kinematic
variables. These are obtained from large samples of simulated events, after
corrections to take into account differences between simulated and real data, as
described in Ref.~\cite{ref25}.

Table 7 shows a summary of backgrounds and efficiencies for separate analyses of
deep-inelastic scattering events (DIS) and  low-multiplicity (LM) events reported
 in Ref.~\cite{ref26}  (DIS events  are defined by the requirement $p^H >$ 1.5 GeV). For all
channels there is good agreement between the observed number of events and the 
  background
prediction. The resulting 90\% confidence level upper limit using the method of
Ref.~\cite{ref22}  is \cite{ref26}
$$P_{\mu\tau} <4.2 \times 10^{-4}~,$$
which corresponds to the exclusion region shown in Fig. 9.

The CHORUS and NOMAD limits can be combined using the method of Ref.~\cite{ref22}.
The combined limit is \cite{ref21}
$$P_{\mu\tau} < 1.3 \times 10^{-4}~.$$

The corresponding exclusion region is outlined in Fig. 9.
\begin{center}
\begin{table}[H]
\caption[]{Summary of NOMAD results \cite{ref26}. $N_\tau$ is the number of expected
$\tau^-$ events for $P_{\mu\tau}$ = 1}
\vskip0.2cm
\begin{tabular}{|l|c|c|c|}
\hline
\multicolumn{1}{|c|}{Analysis}&$N_\tau$&Estimated&Observed\\
&&background&number of\\
&&&events\\
\hline
$\tau^-\to e^-$ DIS&4110&5.3$^{+0.6}_{-0.4}$&5\\
$\tau^-\to h^-(n\pi^0)$ DIS&2232&6.8 $\pm$ 2.1&6\\
$\tau^-\to \rho^-$ DIS&2547&7.3$^{+2.2}_{-1.2}$&8\\
$\tau^-\to \pi^-\pi^-\pi^+$ DIS&1180&6.5 $\pm$ 1.1&5\\
$\tau^-\to e^-$ LM&859&5.4 $\pm$ 0.9&6\\
$\tau^-\to h^-$ LM&357&6.7 $\pm$ 3.0&5\\
$\tau^-\to \rho^-$LM&457&5.2 $\pm$ 2.4&7\\
$\tau^-\to\pi^-\pi^-\pi^+$ LM&108&0.4$^{+0.6}_{-0.4}$&0\\
\hline
\end{tabular}
\end{table}
\end{center}

\section{Long baseline experiments at accelerators}

Long baseline experiments at accelerators extend the sensitivity
 of searches for $\nu_\mu-\nu_x$ oscillations to $\Delta m^2$ as low
as $10^{-3}~{\rm eV}^2$ using $\nu_\mu$ beams of well known properties
which can be monitored and varied if needed. The main goal of these
experiments is to verify that the atmospheric neutrino results \cite{ref6}
  are indeed associated with oscillations, to
establish the nature of the oscillation and to measure its parameters. As
discussed in Ref. \cite{ref6}, the most plausible interpretation of the atmospheric
neutrino results is the occurence of $\nu_\mu-\nu_\tau$ oscillations with $\Delta m^2
\approx 3.5 \times 10^{-3}~{\rm eV}^2$ and full mixing. A $\nu_\mu - \nu_s$ oscillation,
where $\nu_s$ is a new, `sterile' neutrino, is also possible, though less favored.

Table 8 shows a list of parameters for the three existing projects.

\begin{center}
\begin{table}[H]
\caption[]{Long baseline projects}
\vskip0.2cm
\small{\begin{tabular}{|c|c|c|c|c|c|}
\hline
Project&Accelerator&Location of&Distance&$<E_\nu>$&Status\\
&&far detector&(Km)&GeV&\\
\hline
K2K&KEK 12 GeV&Kamioka&250&1.4&Start\\
&proton synchrotron&mine&&&April 1999\\
\hline
NuMI&Fermilab 120 GeV&Soudan&730&16&Start\\
&Main Injector (MI)&mine&&or lower&2002\\
\hline
CNGS&CERN 450 GeV&Gran Sasso&732&30&not yet\\
&SPS&Lab&&or lower&approved\\
\hline
\end{tabular}}
\end{table}
\end{center}

\subsection{K2K}

The K2K project \cite{ref27} uses neutrinos from the decay of $\pi$ and
$K$ mesons produced by the KEK 12 GeV proton synchrotron and aimed at
the SuperKamiokande detector at a distance of 250 km. The beam
consists mainly of $\nu_\mu$, with an average energy of 1.4 GeV. The $\bar{\nu}_\mu$ and
$\nu_e$ contaminations are 4\% and 1\%, respectively. 
The energy is below threshold for $\tau^-$ production $(E_\nu
\approx$ 3.5~GeV), so no search for $\tau^-$ appearance is possible.

The expected distortion of the $\nu_\mu$ flux at 250
km for a $\nu_\mu$ oscillation with $\Delta m^2 = 3.5 \times
10^{-3}~{\rm eV}^2$  can be detected
by comparing the energy distribution of beam-associated muon-like
events in SuperKamiokande with the distribution measured in a similar,
1 Kton water detector located at 300 m from the neutrino source on the
KEK site where no oscillation effects are expected.

The near detector includes a system of scintillating fibres in water,
a lead glass calorimeter and a muon range telescope to monitor and
measure precisely the $\nu_\mu,~\bar{\nu}_\mu$ and $\nu_e$ energy
spectra and space distributions. The 1 Kton water detector will also measure
 the cross-section for $\pi^0$ production in neutral-current
interactions. The knowledge of this
quantity helps understanding if the  disappearance of
atmospheric $\nu_\mu$'s is the result of oscillations to an active or
to a sterile neutrino.

Expected event rates for a run of $10^{20}$ protons on target (3 years, corresponding to an
effective data taking time of 12 months) are given in Table 9. Figure 11 shows the region
of $\nu_\mu-\nu_x$ oscillation parameters which will be excluded if no significant
difference is observed between the near and far detector.

\begin{table}[H]
\begin{center}
\caption{Event rates in the K2K experiment for 10$^{20}$ protons on target.}
\begin{tabular}{|c|c|c|c|}  
\hline
&\multicolumn{2}{|c|}{Near detector}&\\
&Fine-grained&1 Kton water&Super-K\\
\hline
Fiducial mass& 4 tons&21 tons& 22,500 tons\\
\hline
$\nu_\mu$ CC events& 126,700&408,000&345\\
$\nu_\mu$ NC events&44,900&144,000&120\\
$\pi^\circ$-like events&9,200&29,500&25\\
$\nu_e$ CC events&1,250&4,000&4\\
\hline 
\end{tabular}
\end{center}
\end{table}
The K2K experiment has taken the first data between March and June 1999. Interactions of
beam neutrinos in the Super-Kamiokande detector are identified by their timing because the
beam spill is only 1 $\mu$s wide. Clocks at the near and far detector are synchronized with
an accuracy of 100 ns using time signals provided by the Global Positioning System (GPS). During
this running period four events have occurred in a time window of $\pm 50 \mu$s centred on the
arrival time of the neutrino beam pulse. These events are all contained in a 1 $\mu$s wide bin at
$t$ = 0, demonstrating their beam origin. Of these, only one event (with two Cerenkov rings) is a
neutrino interaction in the fiducial volume of the Super-Kamiokande detector.

The K2K experiment has resumed data taking at the end of October 1999.
\begin{figure}[H]
\centering\epsfig{figure=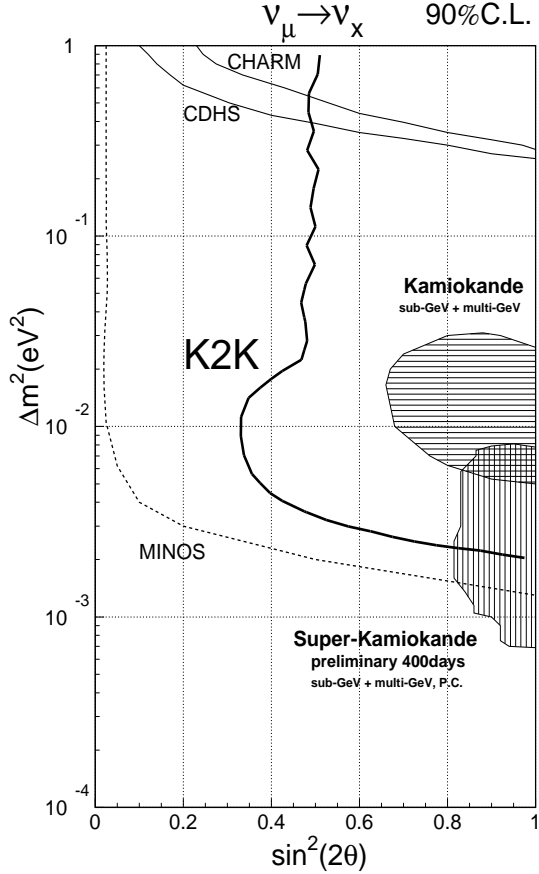,width=7cm}
\caption[]{Region of $\nu_\mu-\nu_\tau$ or $\nu_\mu-\nu_s$ oscillation parameters
excluded at the 90\% confidence level if no oscillation signal is detected by the K2K
experiment after three years of data taking.}
\end{figure}

\subsection{NuMI and the MINOS experiment}

The NuMI project uses neutrinos from the decay of $\pi$ and $K$ mesons
produced by the new Fermilab Main Injector (MI), a 120 GeV proton
synchrotron capable of accelerating $5 \times 10^{13}$ protons with a
cycle time of 1.9 s. The expected number of protons on target is $3.6
\times 10^{20}$/y. The decay pipe is 675 m long.

The neutrino beam will be aimed at the Soudan mine in Minnesota  at a distance of 730 km from the
proton target. The beam will consist primarly of $\nu_\mu$, with 0.6\% $\nu_e$.
Three different neutrino beams have been designed, with average energies
of $\sim$ 3, 8 and 16 GeV corresponding   
 to different tunes and locations of the focusing elements.
For the high energy beam the expected number of events is $\sim$ 3000/y
for a detector mass of 1000 tons.

For an oscillation with
$\Delta m^2 = 3.5 \times 10^{-3}~{\rm eV}^2$ the $\nu_\mu$ flux at 730 km is maximally suppressed at
neutrino energies of 2 GeV. Such a distortion can be best detected using the
lowest energy beam.

The MINOS experiment \cite{ref28} will use two detectors, one (the `near
detector') located at Fermilab, the other (the `far detector') located
in a new underground hall to be built at the Soudan site at a depth of
713 m (2090 m of water equivalent). Both detectors are
iron-scintillator sandwich calorimeters with a toroidal magnetic field
in the iron plates.

The far detector (Fig. 12) has a total mass of 5400 tons and a
fiducial mass of\break\hfill 3300 tons. It consists of magnetized octagonal iron plates, 2.54
cm thick, interleaved with active planes of 4 cm wide, 8 m long
scintillator strips providing both calorimetric and tracking
information. The near detector has a total mass of 920 tons an a fiducial mass of
100 tons. It will be installed 250 m dowstream from the end of the
decay pipe.

\begin{figure}[H]
\begin{center}
\begin{turn}{-90}
\epsfig{figure=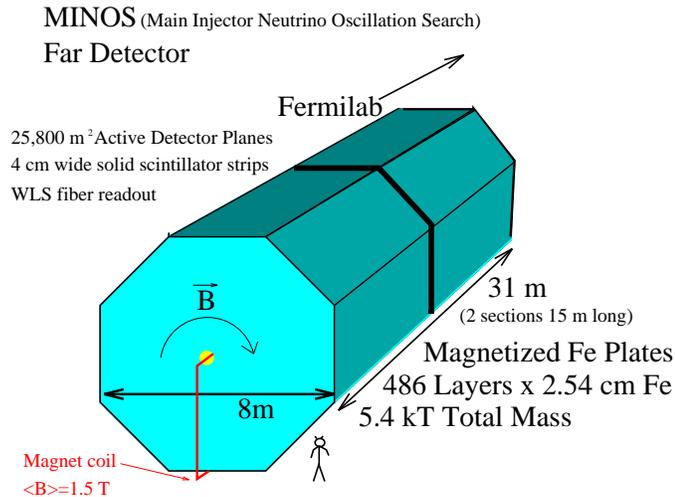,width=6.5cm}
\end{turn}
\caption[]{Sketch of the MINOS detector at the far location.}
\end{center}
\end{figure}

The comparison of $\nu_\mu$ charged-current event rate and energy
distribution in the two detectors will be sensitive to oscillations
which can be detected with a statistical significance of at least four
standard deviations over the full parameter space currently suggested
by the atmospheric neutrino results (see Fig. 13). The oscillation
parameters will be measured precisely over most of this region.

\begin{figure}[H]
\centering\epsfig{figure=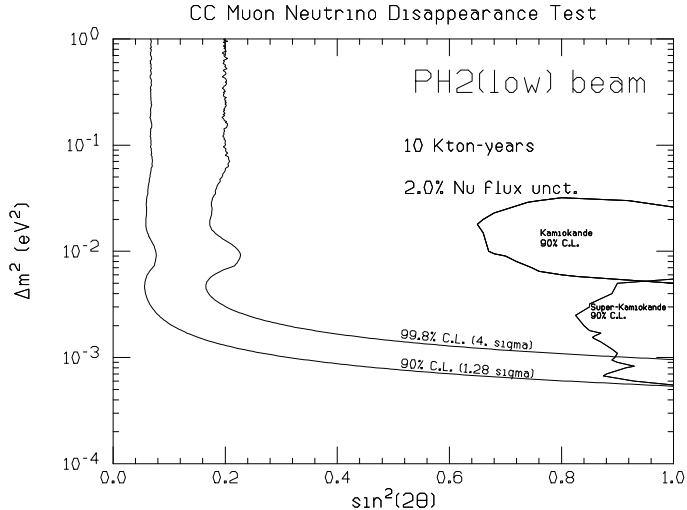,width=9cm}
\caption[]{Excluded region (90\% confidence) and 4$\sigma$ discovery region for
 a\break\hfill 10 kton $\times$ y exposure
of the MINOS experiment from a comparison of the $\nu_\mu$ CC event spectra in the far and near detector.}
\end{figure}

The measurement of the ratio between neutral- and charged-current
event rate (NC/CC) is very important because it will be used to
discriminate between $\nu_\mu-\nu_\tau$ and $\nu_\mu-\nu_s$
oscillations. In the former case NC/CC is larger in the far detector,
while for $\nu_\mu-\nu_s$ oscillations it has the same value in the
near and far detector.

The MINOS experiment should begin data taking at the end of the year
2002.

\subsection{The CNGS project}

The CNGS project (CERN Neutrinos to Gran Sasso) has not yet been approved.
It consists in aiming a neutrino beam from the CERN 450 GeV SPS to the
Gran Sasso National Laboratory in Italy at a distance of 732 km. The
three existing underground halls at Gran Sasso, under $\sim$ 4000 m of
water equivalent, are already oriented towards CERN and
ICARUS \cite{ref29}, a 600 ton detector suitable for oscillation
searches, will start operation in the year 2001 to search for proton
decay and to study atmospheric and solar neutrinos.

If approved before the end of 1999, the CNGS beam will be operational
in the year 2005. It will be used for $\nu_\tau$ appearance
experiments, for which a detector in a `near' location should not be
necessary. The rate of $\nu_\tau$ charged-current interactions from
$\nu_\mu-\nu_\tau$ oscillations is given by
\begin{equation}
N_\tau = A\int\phi_{\nu_\mu}(E)~P_{\mu\tau}(E)~\sigma_\tau(E)~dE
\end{equation}
where $A$ is a normalization constant proportional to the detector
mass, $\phi_{\nu_\mu}(E)$ is the $\nu_\mu$ energy spectrum at the
detector, $P_{\mu\tau}(E)$ is the $\nu_\mu-\nu_\tau$ oscillation
probability and $\sigma_\tau~(E)$ is the cross-section for $\nu_\tau$
charged-current interactions. The integration lower limit is set by
the energy threshold for $\tau^-$ production, 3.5 GeV.

Under the assumption of two-neutrino mixing, $P_{\mu\tau} (E)$ is
given by\break\hfill Eq.~(2). For a large fraction of the $\Delta m^2$ interval
suggested by the atmospheric neutrino results the condition 1.27
$\Delta m^2~L/E <1$ holds for $L$ = 732 km and $E>$ 3.5~GeV. In this
case Eq. (5) can be approximated as
\begin{equation}
N_\tau =  1.61~A~\sin^2 (2\theta)~(\Delta
m^2)^2~L^2~\int~\phi_{\nu_\mu} (E)\sigma_\tau(E)\frac{dE}{E^2}~.
\end{equation}

In this approximation $N_\tau$ varies as $(\Delta m^2)^2$ and
 the neutrino beam must be designed
with the goal of maximizing the integral of  Eq. (6) which does not
depend on $\Delta m^2$. A preliminary beam design, based on a 1000 m
long decay tunnel, is described in Ref.~\cite{ref30}. The design has been
improved \cite{ref31} to optimize the $\tau^-$ production rate. The
$\nu_\mu$ mean energy is 17 GeV and the rate of $\nu_\mu$
charged-current interactions is 2448/y for a detector with a
mass of 1000 tons (this value corresponds to $4.5 \times 10^{19}$
protons on target, which is a realistic figure for a one-year run
after the shut-down of LEP).

The rates of other neutrino events relative to $\nu_\mu$ events are
0.007, 0.02 and 0.0007 for $\nu_e,~\bar{\nu}_\mu$ and $\bar{\nu}_e$,
respectively. Table 10 lists the expected yearly rates of $\tau^-$
events \cite{ref31} for three different $\Delta m^2$ values.

\begin{center}
\begin{table}[H]
\caption[]{Yearly rate of $\tau^-$ events vs $\Delta m^2$ for a 1000 ton
detector \cite{ref31} (1 year =\break\hfill 4.5 $\times 10^{19}$ protons on target).}
\vskip0.2cm
\begin{tabular}{|c|c|}
\hline
$\Delta m^2~({\rm eV}^2)$&$N_\tau$\\
\hline
$10^{-3}$&2.48\\
\hline
$3\times 10^{-3}$&21.7\\
\hline
$5\times 10^{-3}$&58.5\\
\hline
\end{tabular}
\end{table}
\end{center}

ICARUS \cite{ref29} is a new detector concept based on a liquid Argon
Time Projection Chamber (TPC) which allows three-dimensional
reconstruction of events with spatial resolution of the order of 1
mm. A 600 ton ICARUS module is presently being constructed. The cryostat
cold volume (534 m$^3$) is 19.6 m long and 4.2 m high. Three
additional modules will be built if the operation of the first module
is successful.

ICARUS will search for $\tau$ appearance using kinematical criteria
similar to those used in the NOMAD experiment (see Section 4.2).
However, for the $\tau^- \to e^-$ decay channel a background rejection
power in excess of 10$^4$ was needed in NOMAD, while in ICARUS a
rejection power of $\sim~10^2$ is sufficient because of the much
smaller number of events. This allows looser selection
criteria and the detection efficiency for $\tau^- \to e^-$ events
becomes  $\sim$ 50\%. With four modules one expects
$\sim~10~\tau^-$ events/y for a $\nu_\mu-\nu_\tau$ oscillation
with $\Delta m^2 = 5 \times 10^{-3}~{\rm eV}^2$ and full mixing, to be
compared with a background of 0.25 events.

The $\tau$ identification for other $\tau$ decay channels is
presently under study. Since the $\tau^-$ production rate at low
$\Delta m^2$ is proportional to $(\Delta m^2)^2$, a detector
consisting of four ICARUS modules should be sensitive to oscillations
with $\Delta m^2 \ge 2 \times 10^{-3}~{\rm eV}^2$ after a running 
time of four years.

Very recently, a new detector (ICANOE) with a total active mass of 9,300 tons has been proposed
\cite{ref32}. This detector consists of  several ICARUS modules interleaved with conventional
calorimeters made of magnetized iron.

Another interesting detector concept for a $\nu_\tau$ appearance
search is\break\hfill OPERA \cite{ref33}. The detection of one-prong $\tau$ decays
is performed by measuring the $\tau^-$ decay kink in space, as
determined by two track segments measured with very high precision in
nuclear emulsion.

Figure 14 illustrates the OPERA concept. The main component of the
target are 1 mm thick Pb plates where most neutrinos interact.
One-prong $\tau$ decays occurring
 in  the 3 mm gap between the
emulsion detectors ES1, ES2 are expected to result in observable
kinks. ES1 and ES2 consist of two 50 $\mu$m thick emulsion layers
glued to a 100 $\mu$m plastic foil. The 3 mm gap between ES1 and ES2
is filled with a very low density spacer to which ES1 and ES2 are
glued to ensure that their relative positions are stable and precisely
known.
\begin{figure}[H]
\centering\epsfig{figure=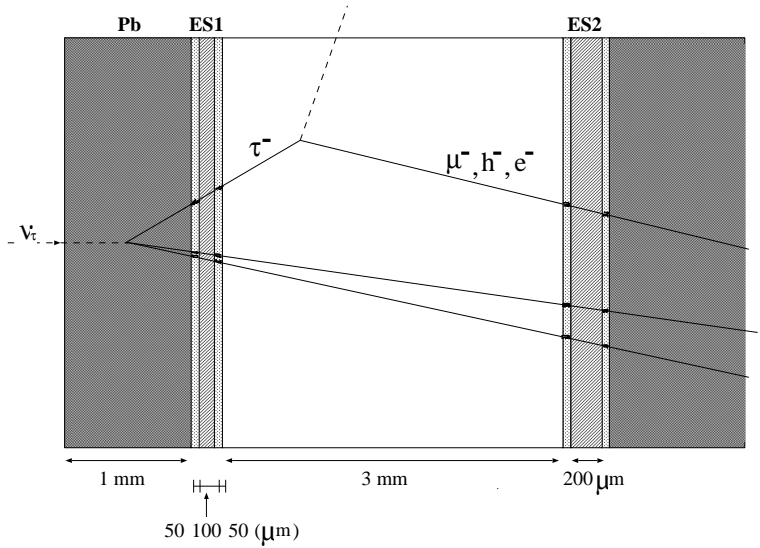,width=10cm}
\caption[]{Schematic structure of the OPERA target.}
\end{figure}

The global $\tau^-$ detection efficiency is estimated to vary between
0.29 at low $\Delta m^2$ and 0.33 at $\Delta m^2 \approx 10^{-2}~{\rm
eV}^2$. For a   run of one year (4.5 $\times 10^{19}$ protons on target)
the background is expected to be 0.4 events, mostly from charm
production and decay in events in which the primary $\mu^-$ was not
identified. The number of expected $\tau^-$ events for full mixing is
listed in Table 11 for different $\Delta m^2$ values.

\begin{center}
\begin{table}[H]
\caption[]{Number of detected $\tau^-$ in OPERA per 4.5 $\times~10^{19}$
protons on target\break\hfill (1 year)  (from Ref. \cite{ref33})}
\vskip0.2cm
\begin{tabular}{|c|c|}
\hline
$\Delta m^2~({\rm eV}^2)$& Detected $N^-/y$\\
\hline
$1.0 \times 10^{-3}$&0.5\\
\hline
$3\times 10^{-3}$&4.5\\
\hline
$5.0\times 10^{-3}$&12.0\\
\hline
\end{tabular}
\end{table}
\end{center}

\section{Searches for new particles using neutrino beams}

The NuTeV (E--815) experiment at Fermilab has searched for a hypothetical Neutral Heavy
Lepton, $L_\mu$, which mixes with $\nu_\mu$. If such a particle exists, it is produced in
meson decays such as $K^+\to \mu^+ L_\mu,~D^+\to\mu^+ L_\mu$ or $D^- \to \mu^- \bar{L}_\mu$.
It would then be present in the high-energy neutrino beam from the Tevatron.

$L_\mu$ is expected to decay to $\mu^-\mu^+\nu_\mu,~\mu^-e^+\nu_e,~\mu^-\pi^+$ or
$\mu^-\rho^+$. In the NuTeV experiment these decays would result in two-track events with a
$\mu^-$ and a positive track originating in a 34 m long region in front of the detector.
Backgrounds from ordinary neutrino interactions in this region are reduced by the use of
Helium bags.

No event consistent with $L_\mu$ decay has been observed \cite{ref34}. Figure 15 shows the
90\% confidence level upper limit on the mixing parameter $U^2$ as a function of the $L_\mu$
mass.

\begin{figure}[H]
\begin{center}
\begin{turn}{90}
\epsfig{figure=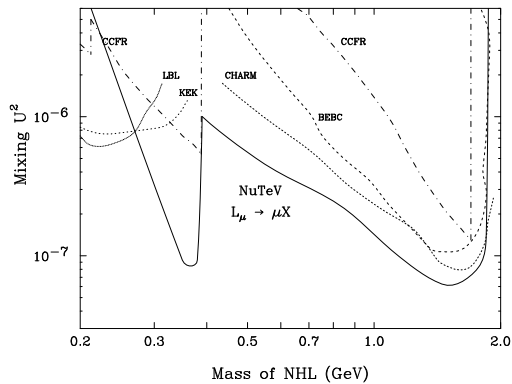,width=5cm}
\end{turn}
\caption[]{NuTeV 90\% confidence level upper limit for $U^2$, the mixing of Neutral Heavy
Leptons (NHL) to $\nu_\mu$, as a function of NHL mass (solid line). Also shown are the
results from previous experiments.}
\end{center}
\end{figure}

\section{Conclusions}

Neutrino experiments at reactors and accelerators have provided so far a wealth of physics
results and will continue to do so in the future. Over the next decade these experiments
will focus on oscillation studies. They will verify that the atmospheric neutrino results
are indeed associated with oscillations and will provide more precise measurements of the
oscillation parameters. They will also verify the large mixing angle MSW solution of the
solar neutrino problem and provide information on the existence (or non-existence) of the
fourth, sterile neutrino which is required if three independent $\Delta m^2$ values are
needed to explain all the observed oscillation signals.

\vspace*{-0.3cm}
\section*{Acknowledgements}

I thank Dario Autiero, Janet Conrad, Guido Drexlin, Klaus Eitel
and Bill Louis for the help provided in preparing this
report.

\end{document}